\RequirePackage{fix-cm}
\documentclass[10pt]{llncs}

\usepackage[T1]{fontenc}
\usepackage{times} 
\usepackage{dcolumn}
\usepackage{endnotes}
\usepackage{graphics}
\usepackage{times}
\usepackage{epsfig}
\usepackage{graphicx} 
\usepackage{textcomp}
\usepackage{amssymb,amsmath}
\usepackage{balance}
\usepackage{url}
\usepackage{listings}
\usepackage{flushend}
\usepackage{color}

\begin{document}


\title{Autoscaling Bloom Filter
}
\subtitle{Controlling Trade-off Between True and
False Positives}

\author{Denis Kleyko \inst{1}      \and
        Abbas Rahimi \inst{2}   \and 
	Ross~W.~Gayler\inst{3}   \and
        Evgeny Osipov \inst{1}
}

\institute{$^1$ Lule\aa{} University of Technology, Lule\aa{}, Sweden.\\
\email{\{denis.kleyko, evgeny.osipov\}@ltu.se}\\
$^2$ETH Zurich, Zurich, Switzerland.\\
\email{abbas@ee.ethz.ch}\\
$^3$Independent Researcher, Melbourne, Australia.\\
\email{r.gayler@gmail.com} 
}


\maketitle

\begin{abstract}
A Bloom filter is a simple data  structure supporting membership queries on a
set. The standard Bloom filter does  not support the  delete operation,
therefore, many applications use a counting  Bloom filter  to enable deletion.
This paper proposes a generalization of the counting Bloom filter approach,
called ``autoscaling Bloom filters'', which allows adjustment of its
capacity with probabilistic bounds on false positives and true positives. In
essence, the autoscaling Bloom filter is a binarized counting Bloom filter with an
adjustable binarization threshold. We present the mathematical analysis
of the performance as well as give a procedure for minimization of the false
positive rate.

\end{abstract}


\section{Introduction}
\label{sect:intro}



Many applications require fast and memory efficient querying of an item's
membership in a set.  A Bloom filter (BF) is a simple binary data structure,
which supports approximate set membership queries.

The standard BF (SBF) allows adding new elements to the filter and is
characterized by 
a perfect true positive rate (i.e. 1), but
nonzero false positive rate. 
The false positive rate depends on the number of elements to be stored in the filter
and on the filter's parameters, including the number of hash functions and 
the size of the filter.
However, SBF
lacks the functionality of deleting an element. Therefore, a
counting Bloom filter (CBF) \cite{COUNTBF}, providing the delete operation,   is
commonly used. 
When the size of the CBF and the number of
elements to be stored are known, the
number of hash functions can be optimized to
minimize the false positive rate. 

Another practical issue is that the parameters of a BF 
(size of filter and number of hash functions)
can not be altered once it is constructed.
If the current filter does not satisfy 
the performance requirements
 (e.g. false positive rate)
it is necessary
to rebuild the entire filter, which is computationally expensive.
Therefore, the optimization of a BF is problematic and costly when the number of elements to be stored is unknown or varies dynamically. 

To address the
issue of optimizing BF performance without rebuilding the filter,
we propose the autoscaling Bloom filter (ABF), 
which is derived from a CBF 
and allows minimization of the false positive rate 
in response to changes in the number of stored elements
without requiring rebuilding of the entire filter. 
The reduction in false positive rate is achieved by 
optimizing a threshold parameter used to derive the ABF from the CBF. 
ABF operates 
with fixed resources 
(i.e. fixed size storage array and fixed $k$ hash functions) 
for a wide dynamic range of number of input elements to be stored.
The trade-off made by ABF for this flexibility is a slight reduction of the true positive rate
(which is always 1 in CBF). 
It is important to note that a less than perfect true positive rate can be tolerated in many applications including networking \cite{RBF}, and generally in the area of approximate computing where errors and approximations are acceptable 
as long as the outcomes have a well-defined statistical behavior \cite{RESBF}. 
To the best of our knowledge, ABF is a novel simple construction of BFs, 
which makes them particularly useful in scenarios 
where a reduced true positive rate can be tolerated and
where the number of the stored elements is unknown or changes dynamically with time.

\begin{figure}[tb]
\centering
\includegraphics[width=0.7\columnwidth]{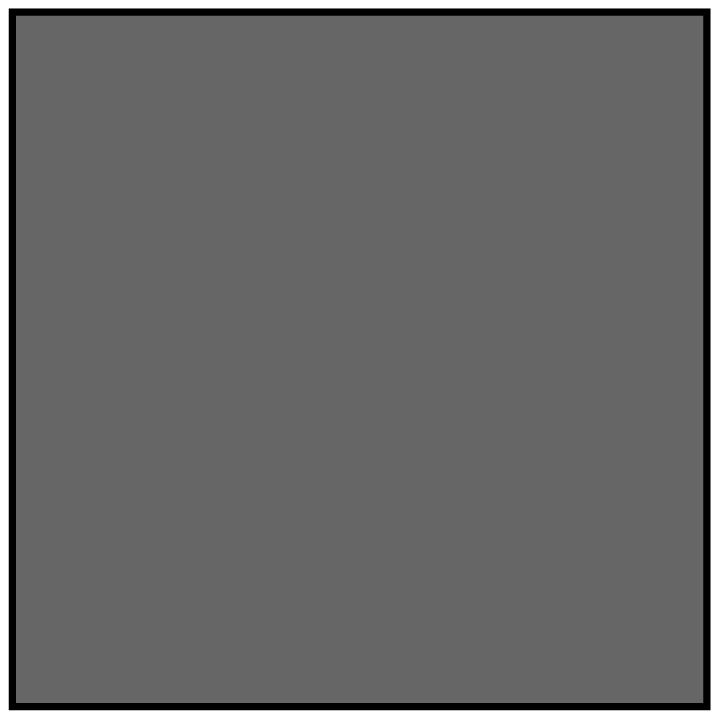}
\caption{An example of the CBF (a), the SBF (b), and two versions of the ABF (c) and (d).}
\label{fig:example}
\end{figure}


ABF belongs to a class of binary BFs and 
is constructed by binarization of  
a CBF with the binarization threshold ($\Theta$) as a parameter.
Querying the ABF also uses a decison threshold ($T$)
to determine whether there is sufficient evidence
to respond that the query item is an element of the stored set.
Both parameters, $\Theta$ and $T$,
can be varied while the ABF is in use
without requiring the filter data structure to be rebuilt.
Figure \ref{fig:example}  
illustrates the main idea behind the ABF. 
Figure
\ref{fig:example}.a 
shows an example
CBF of size 20, which stores four elements
($x_1$ to $x_4$). 
Each element  is mapped to three different positions of the filter;
one position for each of the three hash functions.
The value at each position is the number of elements mapped to that position
by the three hash functions
and varies between 0 and 4 (highlighted by different colors).
The SBF (Figure \ref{fig:example}.b) is formed by setting all nonzero positions
of the CBF to 
one\footnote{
Note that the SBF is a special case of the ABF, 
arising when the binarization threshold is set to zero
}.  
The two lower parts of the figure
present two examples of the ABF with different 
binarization threshold 
($\Theta=1$ and $\Theta=3$ respectively). 
In all four examples, 
the filter
is queried with the unstored element $y$,
testing for membership of the set of stored elements.
The correct answer in every case, obviously, is that $y$ is not a member of the stored set. 
In the SBF example all nonzero positions of $y$ are set to one,
which is interpreted by the SBF algorithm as indicating that the query element
is a member of the stored set,
thus generating a false positive response.
In contrast, in Figure \ref{fig:example}.c,
$y$ has only one position in common with the ABF while all
elements $x_i$ have at least two positions.
Thus, a decision threshold $T$ (for the number of activated positions)
can be chosen such that $y$ will be correctly rejected by the ABF
while all the stored elements are correctly reported as present.
On the other hand, for the ABF in  Figure \ref{fig:example}.d,
the binarization threshold
($\Theta=3$) is too low and it is not possible
to set a decision threshold $T$ such that all stored elements $x_i$
are reported as present.


Mathematically, the ABF has its roots 
in the theory of sparse distributed data representations \cite{Rachkovskij2001}.
ABF can also be 
interpreted
in terms of hyperdimensional computing \cite{Kanerva09},
where everything is represented as high-dimensional vectors
and computation is implemented by arithmetic operations on the vectors.
Both sparse distributed representations  and hyperdimensional computing
can be conceptualised as weightless artificial neural networks. 
From a neural processing point of view, 
BFs are a special case of an artificial neural network 
with two layers (input and output),
where each position in a filter is implemented as a binary neuron. 
Such a network does not have interneuronal connections.
That is, output neurons (positions of the filter) 
have only individual connections with themselves 
and the corresponding input neurons.

This paper explores a direct correspondence between BFs and hyperdimensional representations. 
BFs are treated as a special case application of distributed representations
where each element stored in the BF 
is represented as a hyperdimensional binary vector 
constructed by the  hash functions.
The mathematics of sparse hyperdimensional computing \cite{Rachkovskij2001} (SHC) 
is used for describing the behavior of the proposed ABF. 
The construction of the filter itself 
corresponds to the bundling operation \cite{Rachkovskij2001} of binary vectors.  

The main contributions of the paper are as follows:
\begin{itemize}
  \item It proposes the ABF, which is a generalization of the CBF 
  with probabilistic bounds on false positives and true positives;   
  \item It presents the mathematical analysis and experimental evaluation 
  of the ABF properties;  
  \item It 
  gives 
 a procedure for {\it automatic} minimization 
  of the false positive rate 
  adapting
  to the number of the elements 
stored
in the filter.
\end{itemize}

The paper is structured as follows: Section \ref{sect:related}  presents a
concise survey of the related approaches. Section \ref{sect:math}  describes the
ABF and introduces analytical expressions characterizing its performance.  The
evaluation of the ABF is presented in Section \ref{sect:eval}.  The paper is
concluded in Section \ref{sect:conclusions}.

\section{Related Work}
\label{sect:related}

A recent probabilistic analysis of the SBF is presented in \cite{PABF18}. Detailed surveys on BFs and their applications are provided in \cite{TPBF} and \cite{SUR04}. 
Recent applications of BFs and their modifications include certificate revocation for smart grids \cite{BFSG}.  
An important aspect for the applicability of BFs 
in modern 
networking applications is
the
processing speed of a filter. 
In order to improve the speed of the membership 
check, the authors 
in \cite{UFBF} proposed a novel
filter 
type called Ultra-Fast BFs. In \cite{BFAES} it was shown that BFs can be accelerated 
(in terms of processing speed) by using particular types of hashing functions.

This section overviews the approaches most relevant to the presented ABF approach. 
One direction of research is to propose new types of data structures supporting approximate membership queries.
For example, recently proposed invertible Bloom lookup tables \cite{IBLT14}, quotient filters \cite{FLASH}, counting quotient filters \cite{GPCF}, TinySet \cite{TINYSET},  and cuckoo filters \cite{CF} support dynamic deletion. Another popular research topic is to improve the performance of the SBF via modifications of the original approach.
The ternary BF
\cite{TBF}  improves the performance of the CBF as it only allows three possible
values of each position. The deletable BF \cite{DBF} uses additional
positions in the filter, which are used to support  the deletion of elements
from the filter without introducing false negatives.  The complement Bloom
Filter \cite{CBF} uses an additional BF in order to  identify the trueness of BF
positives.  The on-off BF \cite{ONOFF} reduces false positives
by including in the filter  additional information 
about those elements that
generate false positives. 
Fingerprint Counting BF \cite{ICBF16} is a modification improving the CBF 
with the usage of fingerprints on the filter elements. 
In \cite{BFSG}, the authors 
propose to use two BFs and an external mechanism in order to resolve cases when the membership is confirmed by both filters. In a similar fashion the cross-checking BF \cite{CCBF} constructs several additional BFs,  which are used to cross-check the main BF if it issues a
positive result.
The scalable Bloom filter \cite{SBF} 
can
maintain the desired false positive rate even when the number of stored elements is unknown. 
However, it has
to maintain a series of BFs in order to do so.
The retouched BF (RBF) \cite{RBF}
is conceptually
the most relevant approach to the ABF since it allows some
false negatives 
as a trade-off for decreasing the
false positive rate.  The major difference to the
proposed approach is that  RBF eliminates 
false positives that are known
in advance. 
When the potential false positives are not known in advance
the RBF could randomly 
erase several nonzero positions of the filter.

 In contrast to the previous work, the ABF is suitable for reducing
the
false positive rate  even when the whole universe of elements is either unknown
or is too large to use additional mechanisms for encoding the elements not
included in the filter.

\section{Autoscaling Bloom Filter}
\label{sect:math}


\subsection{Preliminaries: BFs}

At the initialization phase a BF can be seen as a vector of length $m$ where all positions are set to zero. The value of $m$ determines the size of the filter.
In order to store in the filter an element $q$, from the universe of elements, the element should be mapped into the filter's space. This process is usually seen as application of $k$ different hash functions to the element. The result of each hash function is an integer between 1 and $m$. This value indicates the index of the position of the filter which should be updated. In the case of the SBF, an update corresponds to setting the value of the corresponding position of the SBF to 1. If the position already has value 1 it stays unchanged. In the case of the CBF, an update corresponds to incrementing the value of the corresponding position of the CBF by 1. Thus, when storing a new element in the filter at most $k$ positions of the filter update their values. Note that there is a possibility that two or more hash functions return the same result. In this case, there would be less than $k$ updated positions. However, it is usually recommended 
to choose hash functions such that they have a negligible probability of returning the same index value. 
Therefore, without loss of generality, 
suppose that the $k$ results of $k$ hash functions applied to $q$ never coincide. 
That is,
all $k$ indices pointing to positions in the filter are unique.

Instead of considering the result of mapping $q$ as the $k$ indices produced by the hash functions, it is convenient to represent 
the mapping in the form of the SBF that stores 
the
single element $q$. This SBF is sometimes called the individual BF. 
It is a vector with $m$ positions,
where values of only $k$ positions are set to one, 
and the rest to zero. 
The nonzero positions 
are determined by the hash functions applied to $q$. The representation of an element $q$ in this form is denoted as $\textbf{q}$. Note that throughout this section bold terms denote vectors.
Given this vectorized form of representation, the CBF (denoted as $\textbf{CBF}$) storing a set of $n$ elements $x_i$ can be calculated as the sum of representations (denoted as $\textbf{x}_i$) of each individual element $x_i$ in the set: 
~
\begin{equation}
\textbf{CBF}= \sum_{i=1}^{n} \textbf{x}_i. 
\label{eq:cBF}
 \end{equation}
~
The SBF (denoted as $\textbf{SBF}$) 
representing the set of elements
is related to the CBF 
representing the same set of elements
as follows: 
~
\begin{equation}
\textbf{SBF}= [\textbf{CBF}>0], 
\label{eq:sBF}
 \end{equation}
where [] means 1 if true and 0 otherwise 
(applied elementwise to the argument vector).

Given the values of  $m$ and $n$, 
the value of
$k$ that minimizes 
the
false positive rate 
(see also \cite{BFFPR}, \cite{BFFRP10} for recent improvements)
for the
SBF (CBF) can be found as:
~
\begin{equation}
k=(m/n) \ln 2.
\label{eq:optk}
 \end{equation}

When performing the set membership query operation 
with query element $q$ (represented by $\textbf{q}$)
on an SBF containing $q$,
the dot product ($d$) between
$\textbf{SBF}$ and $\textbf{q}$  
must equal the number of nonzero positions in
$\textbf{q}$,
i.e. $k$: 
~
\begin{equation}
d(\textbf{SBF}, \textbf{q}) = \textbf{SBF} \cdot \textbf{q}=k
\label{eq:dot}
 \end{equation}
~

\subsection{Preliminaries: probability theory}

Two probability distributions are useful for the analysis presented here. These are binomial and hypergeometric distributions. Both are discrete. They describe the probability of $s$ successes (draws for which the drawn entities are defined as successful)  in $g$ random draws from a finite population of size $G$ that contains exactly $S$ successful entities. The difference between binomial and hypergeometric distributions is that the binomial distribution describes the probability of $s$ successes in $g$ draws with replacement while the hypergeometric distribution describes the probability of $s$ successes in $g$ draws without replacement.    

Note that if 1 denotes a successful draw while 0 denotes 
a
failure draw,
then we can represent $g$ draws from a distribution as a binary vector of length $g$. This binary vector corresponds to a realization of a (hypergeometric/binomial) experiment.
The probability of a success in a particular position of the realization for both distributions is: 
~
\begin{equation}
p_s = S/G.
\label{eq:p1dist}
 \end{equation}
~
The difference is that for the binomial distribution positions are independent while for the hypergeometric distribution they are not. For example, if the actual values of some positions are known for the realization of a hypergeometric experiment then the probability of a success for the rest of the positions should be updated accordingly. This is because draws from the population are done without replacement.

If the random variable $Z$ is described by the binomial distribution 
(denoted as $Z \sim \text{B}(g,p_s)$),
then 
the probability of getting exactly $s$ successes in $g$ draws is described by the probability mass function:
~
\begin{equation}
\Pr(Z=s)= \binom{g}{s}p_s^s (1-p_s)^{g-s}. 
\label{eq:pmfbd}
 \end{equation}
~
As the probability mass function for the hypergeometric distribution is not used below it is omitted here. 

\subsection{Preliminaries: relation between BFs and probability theory}

The hypergeometric distribution comes into 
play 
when considering the mapping of an element $q$.
Given the assumption that the results of hash functions do not coincide, the mapping $\textbf{q}$ of an element $q$ is a binary vector of length $m$ with exactly $k$ positions having value 1 and the rest 0. Because hash functions map different elements into different indices, a mapping $\textbf{q}$ can be seen as a single realization of the experiment from the hypergeometric distribution with $g=m$ draws from the finite population of size $G=m$ that contains exactly $S=k$ successes 
(positions set to 1).  
In this case $g=G$.
Therefore, 
the probability of exactly $s=k$ successes is 1 and all other probabilities are 0. The probability of a success in a particular position is:
~
\begin{equation}
p_1=p_s=k/m. 
\label{eq:p1}
\end{equation}
~
A value in $i$th position of $\textbf{CBF}$ (see (\ref{eq:cBF})) can be seen as a discrete random
variable (denoted as $I$) in the  range $I \in \mathbb{Z} | 0 \leq I \leq n$. Because representations $\textbf{x}_i$ stored in $\textbf{CBF}$ are independent realizations of the hypergeometric experiment, $I$ follows the binomial distribution: $I \sim \text{B}(g,p_s)$ where $g=n$, $p_s=p_1$.        


Given the parameters of the binomial distribution, the probability that $I$ takes the value $v$ can be calculated according to (\ref{eq:pmfbd}):
~
\begin{equation}
\Pr(I=v)= \binom{n}{v}p_1^v (1-p_1)^{n-v}. 
\label{eq:PRcBF}
\end{equation}
According to (\ref{eq:PRcBF}),  the probability of an empty position $p_0$ in the CBF
(and also for SBF)
is:
~
\begin{equation}
p_0=\Pr(I=0)=  \left(1-\frac{k}{m} \right)^n. 
\label{eq:PR0}
 \end{equation}
~
It should be noted that the probability of an empty position $p_0$ in the CBF (SBF) when the results of hash functions
can coincide, is:
~
\begin{equation}
p_0 =  \left(1-(1/m) \right)^{kn}. 
\label{eq:PR0col}
 \end{equation}
In fact, (\ref{eq:PR0})  differs from the standard expression (\ref{eq:PR0col})
for $p_0$.  However, both produce different results only for small lengths of
the filter ($m<50$), which are not 
of practical 
importance.

Because each position in $\textbf{CBF}$  can be treated as an independent realization of $I$, the
expected number of positions $l$ with value $v$ equals:
~
\begin{equation}
l(v)= m \Pr(I=v) = m \binom{n}{v}p_1^v (1-p_1)^{n-v}. 
\label{eq:numcBF}
 \end{equation}
~
\subsection{Definition of Autoscaling Bloom Filter}

Given a CBF, the derived ABF is formed  by setting to zero all positions with values
less than or equal to the chosen 
binarization threshold 
$\Theta$;  positions with values
greater than $\Theta$ are set to one:
~
\begin{equation}
\textbf{ABF} = [\textbf{CBF}>\Theta]. 
\label{eq:aBF}
 \end{equation}
Note that when $\Theta=0$,  the ABF is equivalent to the SBF.

In general, the expected dot product (denoted $\bar{d}_x$)  between the ABF
and an element $x$ included in the filter is 
less than or equal to 
$k$.\footnote{ 
It should be noted that the calculation of expected similarity (e.g., dot product) 
between two vectors, one of which may store the other, 
is a general problem formulation in hyperdimensional computing and can be seen as the "detection" type of retrieval (see \cite{Frady17} for details).}
As the binarization threshold $\Theta$ increases,
more of the nonzero positions in the CBF
are mapped to zero values in the corresponding ABF.
This necessarily reduces the dot product
of the ABF vector with the query vector.
Therefore, there is a need for the second parameter  of the ABF, which
determines the lowest value of dot product indicating the presence  of an
element in the filter. 
Denote this 
decision threshold
parameter as $T$ ($0 \leq T \leq k$),  
then an element of the universe $q$ 
is judged to be
a member of the ABF if and only  if the
dot product between $\textbf{ABF}$ and $\textbf{q}$ is greater than 
or equal to
$T$.

\subsection{Probabilistic characterization of the Autoscaling Bloom Filter}

When the 
binarization
threshold $\Theta$ for the ABF is more than zero, the probability  of an empty position in the ABF (denoted as $P_0$) is higher than in the SBF  
because 
some of the 
nonzero positions in the CBF 
are
set to zero. For a given $\Theta$, the expected $P_0$ is
calculated using (\ref{eq:PRcBF}) as follows:
~
\begin{equation}
P_0 = \sum_{v=0}^{\Theta} \Pr(I=v)= 
\sum_{v=0}^{\Theta} \binom{n}{v}p_1^v (1-p_1)^{n-v}.
\label{eq:P0}
 \end{equation}
Then the probability of 
1
in the ABF (denoted as $P_1$) is:
~
\begin{equation}
P_1 = 1- P_0 = 1 - \sum_{v=0}^{\Theta} \binom{n}{v}p_1^v (1-p_1)^{n-v}.
\label{eq:P1}
 \end{equation}
~
The expected dot product $\bar{d}_x$ for an element $x$ included in the ABF is calculated as:
~
\begin{equation}
\bar{d}_x= k-\frac{m}{n}\sum_{v=0}^{\Theta} v \Pr(I=v).
\label{eq:meandx}
 \end{equation}
~
Note that when  $\Theta=0$, $\bar{d}_x(\textbf{ABF}, \textbf{x})=k$ which corresponds to the SBF 
(see (\ref{eq:dot})).  
In other words, the SBF can be seen as 
a 
special case of the ABF. The calculations in (\ref{eq:meandx}) when $\Theta>0$  can be interpreted in the following way. The dot product between $\textbf{SBF}$ and $\textbf{x}$ is $k$. A position in $\textbf{CBF}$ with value $v>0$ contributes 1 to the values of dot products of $v$ stored elements. Thus, if this position is set to zero in the SBF, there will be $v$ elements with the dot product equal to $k-1$ while the dot products for rest of the elements still equal $k$. Then the expected dot product between the filter and an element is decremented by $v/n$. In fact, the number of positions with value $v$ is unknown but it is possible to calculate the probability $Pr(I=v)$ of such position in $\textbf{CBF}$ using (\ref{eq:PRcBF}). Then the expected number of such positions in $\textbf{CBF}$ is determined via (\ref{eq:numcBF}) and equals $mPr(I=v)$. When the ABF suppresses all such positions each of them decrements the expected dot product by $v/n$. Then the total decrement of the expected dot product by the suppressed positions with value $v$ is expected to be $mvPr(I=v)/n$. 
Because the ABF suppresses all positions with values less than or equal 
to $\Theta$,
the decrements of the expected dot product introduced by each value $v$ should be summed up. 

The expected dot product 
(denoted $\bar{d}_y$)
 between  the ABF and an
element $y$ which is not included in the filter is determined by the number of nonzero positions in the filter and calculated as:
~
\begin{equation}
\bar{d}_y= kP_1.
\label{eq:meandy}
 \end{equation}
~
Both dot products $d_x$ and $d_y$ are  characterized by discrete random
variables (denoted as $X$ and $Y$ respectively)  which in turn are described by 
binomial distributions:
$X \sim \text{B}(k,p_x)$ and   $Y \sim \text{B}(k,p_y)$.         

The success probabilities ($p_x$ and $p_y$)  of these distributions are
determined from the expected values of dot  product as in (\ref{eq:meandx}) and
(\ref{eq:meandy}):
~
\begin{equation}
p_x=\bar{d}_x/k=1-\frac{m}{nk}\sum_{v=0}^{\Theta} v \Pr(I=v),
\label{eq:px}
 \end{equation}
~
\begin{equation}
p_y=\bar{d}_y/k=P_1.
\label{eq:py}
 \end{equation}
~
\subsection{Performance properties of ABF}

Given
the decision threshold $T$, the true positive rate 
($\text{TPR}$) of the ABF can be calculated using
the probability mass function of $X$ as:
~
\begin{equation}
\text{TPR} =  \sum_{d=T}^{k} \Pr(X=d)= 
 \sum_{d=T}^{k} \binom{k}{d}p_x^d (1-p_x)^{k-d}.
\label{eq:trp}
 \end{equation}
~
Similarly, 
the
false positive  rate ($\text{FPR}$) is calculated using the
probability mass function of $Y$ as:
~
\begin{equation}
\text{FPR} =  \sum_{d=T}^{k} \Pr(Y=d)= 
 \sum_{d=T}^{k} \binom{k}{d}p_y^d (1-p_y)^{k-d}.
\label{eq:frp}
 \end{equation}

\section{Evaluation of ABF}
\label{sect:eval}

\subsection{Optimization of ABF's parameters}

In order to choose the best value of $T$ (or even both $\Theta$ and $T$), an
optimization  criterion is needed.  
It is proposed to optimize the accuracy ($\text{ACC}$) of
the filter.
This is defined 
as the average value of true
positive rate and true negative rate: $\text{ACC}=(\text{TPR}+(1-\text{FPR}))/2$.
Note, that this  definition of accuracy is also known as unweighted average
recall.  
Note also, that the accuracy does not have to be the only choice for the optimization  criterion. The choice of $\text{ACC}$ implies that false positives and false negatives are treated as equally costly. 
However, in a practical application this may not be true. Instead, each of the four possible outcomes (True Positive, False Positive, True Negative, False Negative) will have an associated domain-dependent cost. The designer would then optimize the design parameters so as to minimize the cost in the application scenario. In the absence of a specific application we are forced to use a general performance summary. We have chosen to use accuracy as a general summary because it is simple and well understood.

In addition, an application may specify the lowest acceptable
$\text{TPR}$ (denoted as $L_{\text{TPR}}$).
Then the optimal value of $T$ 
(for fixed $\Theta$)
is found as:

\begin{equation}
T_{opt}= \max_{T} (\text{ACC}| \text{TPR} \geq L_{\text{TPR}}). 
\label{eq:topt}
 \end{equation}

In general, both parameters of the ABF, $\Theta$ and $T$, can be optimized as:

\begin{equation}
\max_{\Theta, T} (\text{ACC}| \text{TPR} \geq L_{\text{TPR}}). 
\label{eq:optfilter}
 \end{equation}

\begin{figure}[tb]
\centering
\includegraphics[width=1.0\columnwidth]{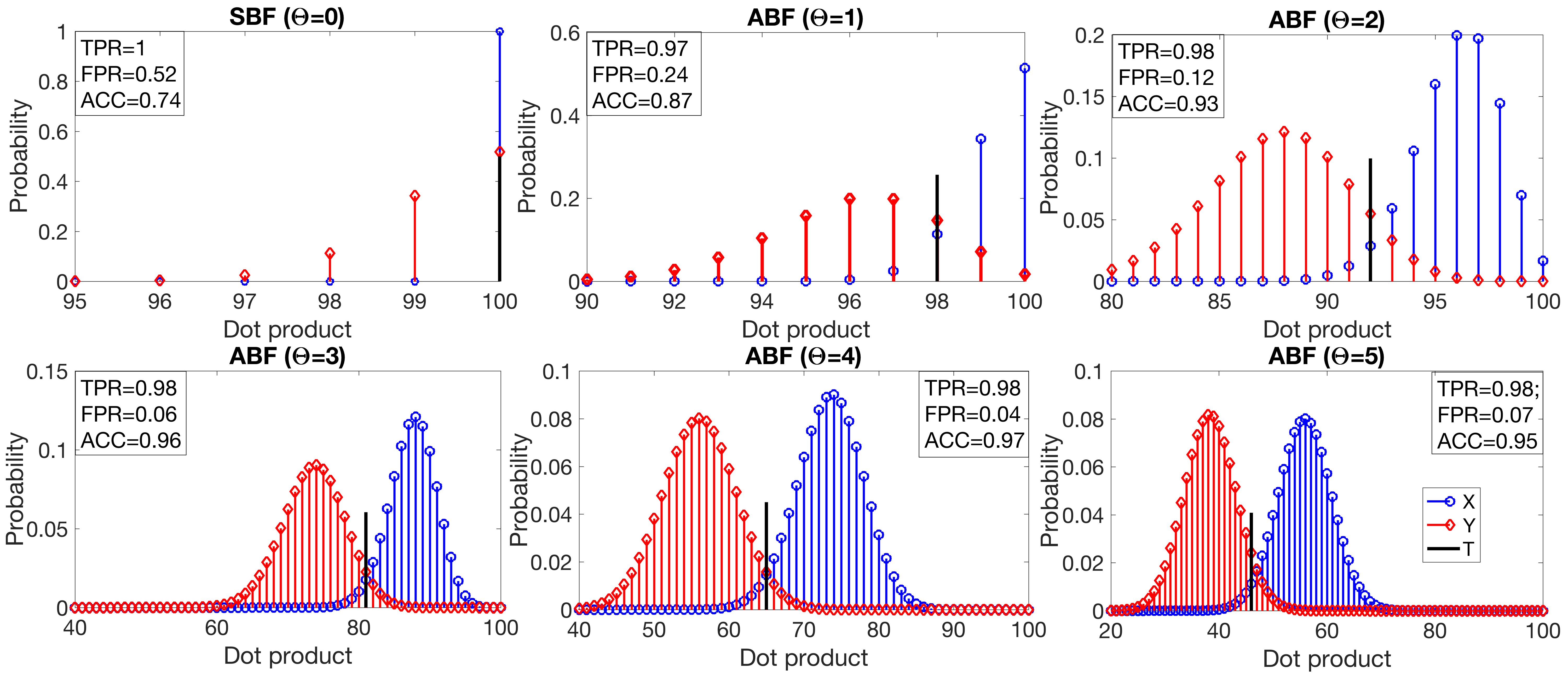}
\caption{Probability mass functions for $X$  (query present)
and $Y$ (query absent) for different  thresholds
$\Theta$ in the range $0 \leq \Theta \leq 5$; $k=100$, $n=500$, $m=10, 000$.}
\label{fig:distr}
\end{figure}

\subsection{An example: ABF in action}

The behavior of ABF  for different $\Theta$ is illustrated in Figure
\ref{fig:distr}. The  length of the CBF (and all derived ABFs) is $m=10, 000$.
It stores $n=500$ unique  elements, each element is mapped to an individual BF
with $k=100$ nonzero positions.  
Note that, the value of $k$ 
in this example
is intentionally
not optimized for the given $m$ and $n$.   
The particular value of $k$ is chosen for demonstration 
purposes
to clearly illustrate the situation when 
the SBF has a high false positive rate
which can be significantly decreased by the ABF. 
Similar effects can be seen
for other values of $k$.

Six ABFs are formed from the CBF using
different thresholds in the range $0 \leq \Theta \leq 5$.  Each plot in Figure
\ref{fig:distr} corresponds to one ABF and depicts probability  mass functions
for $X$ (circle markers) and $Y$ (diamond markers).  
where $X$ and $Y$ denote random variables characterizing distributions of dot products for elements stored in the filter ($X$) and elements not included in the filter ($Y$).

The plot for $\Theta=0$
corresponds to the SBF. In this case,  $X$ is  deterministic and located at
$k=100$ as expected 
given 
$k=100$ nonzero positions  
for the SBF.
Hence,  the optimal value of $T$ is trivially equal to $k$
and $\text{TPR}=100\%$.  
A large portion of the distribution for $Y$  
is also concentrated at $k=100$,
which leads to high $\text{FPR}=52\%$. 
On the other hand, 
the ABFs with $\Theta>0$  
have better separation
of the two distributions.
Much lower \text{FPR} can be achieved 
by reducing the $\text{TPR}$ below 100\%.
The optimal values of $T$ (indicated by black vertical
bars) were found for each value of $\Theta$ according  to (\ref{eq:topt}). The
lowest acceptable  value of $\text{TPR}$, $L_{\text{TPR}}$ was set to 0.97. 
This particular value was chosen to demonstrate 
that, in principle, 
a large reduction of the FPR can be achieved via a small reduction in the TPR. 
The
best values of $\text{TPR}$, $\text{FPR}$, and $\text{ACC}$ for  each plot are
depicted in the figure. 
For example, even changing $\Theta$  from 0 to 1 allows
$\text{FPR}$ to be reduced from $0.52$ to $0.24$ 
at the cost of reducing $\text{TPR}$ by only 3\%.
Overall, the accuracy is improved by 0.13.  The best performance
among the considered range is achieved for $\Theta=4$,  resulting in
$\text{TPR}=0.98$, $\text{FPR}=0.04$, $\text{ACC}=0.97$,  thus, improving the
accuracy of the SBF by 31\%.

\begin{figure}[tb]
\centering
\includegraphics[width=1.0\columnwidth]{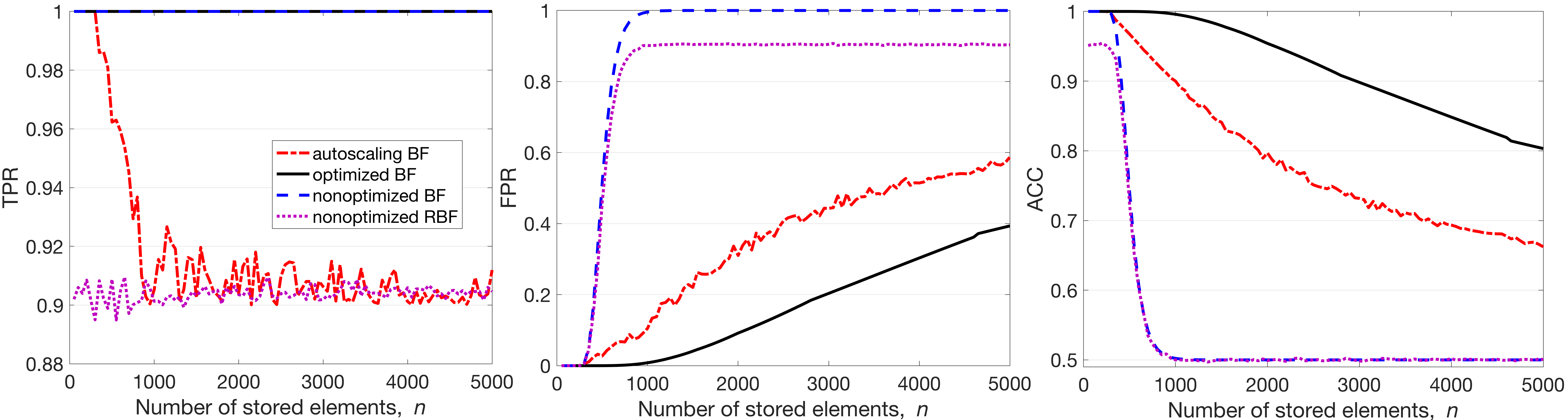}
\caption{
Comparison of performance ($\text{TPR}$,  $\text{FPR}$, and $\text{ACC}$) of
four different BFs against varying number of stored elements $n$ ($50 \leq n \leq 5, 000$ step 50).}
\label{fig:comparison}
\end{figure}

\subsection{Comparison with the optimized BF}

Figure \ref{fig:comparison}  demonstrates the results of comparison of four
filters: the autoscaling BF (dash-dot line), the optimized BF (solid line),
the nonoptimized  BF (dashed line), and the nonoptimized  RBF (dotted line). The nonoptimized  RBF was created via randomly erasing 0.1\% of nonzero positions in the nonoptimized BF.
 Each panel in Figure \ref{fig:comparison}
corresponds to a performance metric:  left - $\text{TPR}$;  center -
$\text{FPR}$; right - $\text{ACC}$.  The performance was studied 
for a range of numbers of unique  elements
 stored in the filter ($50 \leq n \leq 5, 000$). The
length of the filters was the same as in Figure \ref{fig:distr}, $m=10, 000$.  For
the optimized BF, $k$ was calculated as in (\ref{eq:optk}) for each value of $n$ and varied between 1 and 139.
For three other BFs $k$ was fixed to 100.  The ABF was formed
from the CBF according to (\ref{eq:aBF}). 
Only two parameters ($\Theta$ and $T$) of ABF were optimized for each value of $n$ according to (\ref{eq:optfilter}) with $L_{\text{TPR}}=0.9$.
Note that these two parameters do not change 
the hardware resources required for an ABF implementation
since $k$ and $m$ are fixed, while an optimized BF implementation might require
40\% more hash functions. 
This overhead directly translates to a larger silicon area or slower speed for the hardware implementation of the optimized BF compared to the ABF.    

The $\text{TPR}$ of the  optimized and nonoptimized BFs is always 1, while for the ABF and nonoptimized RBF it can be less.
In particular, the $\text{TPR}$ of the ABF varies  in the allowed range between $L_{\text{TPR}}$
and 1.
For large values of $n$ (>1000) the $\text{TPR}$ of the ABF is approximately equal to $L_{\text{TPR}}$. 
In the case of nonoptimized RBF the $\text{TPR}$ was around 0.9 over the whole range of $n$.
The $\text{FPR}$ of all the filters grows with increasing $n$.
As anticipated, the nonoptimized BF soon
(at $n \approx 1000$) achieves $\text{FPR}=1$, 
and stays there until the end. 
A similar behavior is demonstrated by the nonoptimized RBF with the exception that the highest value of $\text{FPR}$ is $0.9$. 
Note that 
with RBF,
the price one has to pay for the lower $\text{FPR}$ is the decreased $\text{TPR}$.
Two other filters, the ABF and the optimized BF, 
demonstrate a smooth increase in $\text{FPR}$.
The $\text{FPR}$
is lower than 1 for both filters even when $n=5,000$
(approximately 0.6 and 0.4 respectively).
The  accuracy curves aggregate the behavior for $\text{TPR}$ and $\text{FPR}$.
For most values of $n$, the  nonoptimized BF and RBF
reach
$\text{ACC}=0.5$ as their $\text{FPR}$s reach the maximal values. 
Their accuracies for large values of $n$ are the same 
because the gain in $\text{FPR}$ 
equals the loss in $\text{TPR}$ for the  nonoptimized RBF.   
The  accuracies of the ABF and the optimized BF smoothly decay
with the growth of $n$, 
being $0.66$ and $0.8$ when $n=5, 000$. 
Thus, the ABF
significantly outperforms the nonoptimized BF and RBF  when their $\text{FPR}$s are
increasing. In general, the performance of the  ABF follows that of the optimized
BF with some constant loss. 
The increase in accuracy from ABF to optimized BF
can be understood as the value delivered
by being able to specify in advance
precisely the number of elements to be stored
in the filter.
The best trade-off between $\text{TPR}$ and
$\text{FPR}$ is in the region of $n$ where $\text{FPR}$ of the nonoptimized BF
is steeply  increasing from 0 to 1.  The important advantage of the ABF over the
optimized BF  is that it does not require the recalculation of the whole filter
as the number  of the stored elements is increasing,
while the optimized BF
must be rebuilt if a new value of $k$ is chosen. 
For example, in the experiments in Figure \ref{fig:comparison} the optimized BF was rebuilt 23 times.  
Furthermore, if the BF is implemented in hardware
then the number of hash functions is most likely fixed by the implementation.



\section{Conclusion}
\label{sect:conclusions}

This paper  introduced the autoscaling Bloom filter. 
The ABF is a generalization
of the standard binary 
BF, derived from the counting BF,
with procedures for achieving
probabilistic bounds on false positives and true positives.
 It was shown that the
ABF can significantly decrease the false positive rate at a
cost of allowing a nonzero false negative rate. 
The evaluation revealed that
the accuracy of the  ABF follows the standard BF  with the optimized number of hash functions with some constant loss. 
As opposed to the optimized BF, the ABF provides means for optimization 
of the filter's performance 
without requiring the entire filter to be rebuilt
when the number of stored elements in the filter is changing dynamically.
This optimization can be achieved while the number of hash functions remains fixed.





\bibliographystyle{unsrt}
\bibliography{bica}

\begin{thebibliography}{10}

\bibitem{COUNTBF}
L.~Fan, P.~Cao, J.~Almeida, and A.~Broder.
\newblock {Summary cache: A scalable wide-area web cache sharing protocol}.
\newblock {\em IEEE/ACM Transaction on Networking}, 8(3):281--293, 2000.

\bibitem{RBF}
B.~Donnet, B.~Baynat, and T.~Friedman.
\newblock {Retouched Bloom filters: Allowing networked applications to trade
  off selected false positives against false negatives}.
\newblock In {\em {ACM CoNEXT conference}}, pages 1--12, 2006.

\bibitem{RESBF}
V.~Akhlaghi, A.~Rahimi, and R.~K. Gupta.
\newblock {Resistive Bloom Filters: From Approximate Membership to Approximate
  Computing with Bounded Errors}.
\newblock In {\em {Conference on Design, Automation and Test in Europe
  (DATE)}}, pages 1--4, 2016.

\bibitem{Rachkovskij2001}
D.~A. Rachkovskij.
\newblock {Representation and Processing of Structures with Binary Sparse
  Distributed Codes}.
\newblock {\em Knowledge and Data Engineering, IEEE Transactions on},
  3(2):261--276, 2001.

\bibitem{Kanerva09}
P.~Kanerva.
\newblock Hyperdimensional computing: An introduction to computing in
  distributed representation with high-dimensional random vectors.
\newblock {\em Cognitive Computation}, 1(2):139--159, 2009.

\bibitem{PABF18}
F.~Grandi.
\newblock {On the analysis of Bloom filters}.
\newblock {\em Information Processing Letters}, 129:35 -- 39, 2018.

\bibitem{TPBF}
S.~Tarkoma, C.~E. Rothenberg, and E.~Lagerspetz.
\newblock {Theory and Practice of Bloom Filters for Distributed Systems}.
\newblock {\em IEEE Communications Surveys and Tutorials}, 14(1):131--155,
  2012.

\bibitem{SUR04}
A.~Broder and M.~Mitzenmacher.
\newblock {Network applications of Bloom filters: A survey}.
\newblock {\em Internet mathematics}, 1(4):485--509, 2004.

\bibitem{BFSG}
K.~Rabieh, M.M.E.A. Mahmoud, K.~Akkaya, and S.~Tonyali.
\newblock {Scalable Certificate Revocation Schemes for Smart Grid AMI Networks
  Using Bloom Filters}.
\newblock {\em IEEE Transactions on Dependable and Secure Computing},
  14(4):420--432, 2017.

\bibitem{UFBF}
J.~Lu, Y.~Wan, Y.~Li, C.~Zhang, H.~Dai, Y.~Wang, G.~Zhang, and B.~Liu.
\newblock {Ultra-Fast Bloom Filters using SIMD techniques}.
\newblock In {\em {2017 IEEE/ACM 25th International Symposium on Quality of
  Service (IWQoS)}}, pages 1--6, 2017.

\bibitem{BFAES}
Y.~Zhang, Z.~Zheng, and X.~Zhang.
\newblock {Efficient Bloom Filter for Network Protocols Using AES Instruction
  Set}.
\newblock {\em IET Communications}, 11(11):1815--1821, 2017.

\bibitem{IBLT14}
S.~Pontarelli, P.~Reviriego, and M.~Mitzenmacher.
\newblock {Improving the performance of Invertible Bloom Lookup Tables}.
\newblock {\em Information Processing Letters}, 114(4):185 -- 191, 2014.

\bibitem{FLASH}
M.A. Bender, M.~Farach-Colton, R.~Johnson, R.~Kraner, B.C. Kuszmaul,
  D.~Medjedovic, P.~Montes, P.~Shetty, R.~P. Spillane, and E.~Zadok.
\newblock {Don't thrash: How to cache your hash on flash}.
\newblock {\em Proceedings of the VLDB Endowment}, 5(11):1627--1637, 2012.

\bibitem{GPCF}
P.~Pandey, M.A. Bender, R.~Johnson, and R.~Patro.
\newblock {A General-Purpose Counting Filter: Making Every Bit Count}.
\newblock In {\em {SIGMOD'17 Proceedings of the 2017 ACM International
  Conference on Management of Data}}, pages 775--787, 2017.

\bibitem{TINYSET}
G.~Einziger and R.~Friedman.
\newblock {TinySet - An Access Efficient Self Adjusting Bloom Filter
  Construction}.
\newblock {\em IEEE/ACM Transaction on Networking}, 25(4):2295--2307, 2017.

\bibitem{CF}
B.~Fan, D.G. Andersen, M.~Kaminsky, and M.D. Mitzenmacher.
\newblock {Cuckoo Filter: Practically Better Than Bloom}.
\newblock In {\em {CoNEXT'14 Proceedings of the 10th ACM International on
  Conference on emerging Networking Experiments and Technologies}}, pages
  75--88, 2014.

\bibitem{TBF}
H.~Lim, J.~Lee, H.~Byun, and C.~Yim.
\newblock {Ternary Bloom Filter Replacing Counting Bloom Filter}.
\newblock {\em IEEE Communications Letters}, 21(2):278--281, 2017.

\bibitem{DBF}
C.~E. Rothenberg, C.~A.~B. Macapuna, F.~L. Verdi, and M.~F. Magalhaes.
\newblock {The deletable Bloom filter: A new member of the Bloom family}.
\newblock {\em IEEE Communications Letters}, 14(6):557--559, 2010.

\bibitem{CBF}
H.~Lim, J.~Lee, and C.~Yim.
\newblock {Complement Bloom Filter for Identifying True Positiveness of a Bloom
  Filter}.
\newblock {\em IEEE Communications Letters}, 19(11):1905--1908, 2015.

\bibitem{ONOFF}
L.~Carrea, A.~Vernitski, and M.~Reed.
\newblock {Yes-no Bloom filter: A way of representing sets with fewer false
  positives}.
\newblock {\em ArXiv:1603.01060}, pages 1--28, 2016.

\bibitem{ICBF16}
S.~Pontarelli, P.~Reviriego, and J.A. Maestro.
\newblock {Improving counting Bloom filter performance with fingerprints}.
\newblock {\em Information Processing Letters}, 116(4):304 -- 309, 2016.

\bibitem{CCBF}
H.~Lim, N.~Lee, J.~Lee, and C.~Yim.
\newblock {Reducing false positives of a Bloom filter using cross-checking
  Bloom filters}.
\newblock {\em Applied Mathematics and Information Sciences}, 8(4):1865--1877,
  2014.

\bibitem{SBF}
P.S. Almeida, C.~Baquero, N.~Preguica, and D.Hutchison.
\newblock {Scalable Bloom Filters}.
\newblock {\em Information Processing Letters}, 101(6):255--261, 2007.

\bibitem{BFFPR}
P.~Bose, H.~Guo, E.~Kranakis, A.~Maheshwari, P.~Morin, J.~Morrison, M.~Smid,
  and Y.~Tang.
\newblock {On the false-positive rate of Bloom filters}.
\newblock {\em Information Processing Letters}, 108(4):210--213, 2008.

\bibitem{BFFRP10}
K.~Christensen, A.~Roginsky, and M.~Jimeno.
\newblock {A new analysis of the false positive rate of a Bloom filter}.
\newblock {\em Information Processing Letters}, 110(21):944 -- 949, 2010.

\bibitem{Frady17}
E.~P. Frady, D.~Kleyko, and F.~T. Sommer.
\newblock Theory of the superposition principle for randomized connectionist
  representations in neural networks.
\newblock {\em arXiv:1707.01429}, pages 1--42, 2017.

\end{thebibliography}

%
%

\end{document}